# From Non-scattering to Super-scattering with the Topology of Light and Matter

Hooman Barati Sedeh, Natalia M. Litchinitser [*]

Department of Electrical and Computer Engineering, Duke University, 27708 Durham, NC, USA.

[*] Corresponding author: natalia.litchinitser@duke.edu

**Abstract**

**Electric anapole states, arising due to the destructive interferences of primitive and toroidal electric dipole moments, have been recently introduced as the fundamental class of non-scattering sources with several potential applications ranging from nonlinear optics to thermodynamics thanks to their field confinement and minimal scattering fingerprints. Nevertheless, other non-radiating sources are also possible and can provide significant energy confinement at the nanoscale if they spectrally overlap with their electrical counterparts. On the other hand, super-scattering states represent the opposite regime of light-matter interaction wherein the scattering cross-section of a particular multipolar moment exceeds the single-channel limit, leading to a strong scattering in the direction of the pump beam. Here, we demonstrate that the interplay between the topology of incident light and the subwavelength scatterer can lead to these two opposite regimes of light-matter interactions within an isolated all-dielectric meta-atom. In particular, we show that the presented scatterer can support new non-scattering states, called hybrid anapole, leading to significant suppression of the far-field radiation and enhancement of electromagnetic energy inside the meta-atom. We also explore the role of particle orientation and its inversion symmetry on the scattering response and show how switching between non-scattering to super-scattering states can occur within the same platform. The presented study elucidates the role of light and matter topologies in the scattering response of subwavelength meta-atoms and reveals the formation of two extreme opposite regimes of light-matter interaction, opening new avenues in several applications ranging from nonlinear optics to spectroscopy.**

▪ **Introduction**

Before the proposal of Bohr's model, in the early days of electron discovery, several models, such as non-radiating sources that do not radiate energy into far-field, had been considered to explain the atomic structure and address the paradox of unstable atoms, which originated from Rutherford's attempt to describe the electron's motion based on the dynamics of planets in the solar system [1]. While Bohr eventually addressed this problem, the field of non-radiating sources continued to be implemented in other branches of science, such as the physics of elementary particles, quantum field theory, nuclear, and dark-matter physics [2]. However, despite the rich history and successful progress of this field of research, the direct evidence of optical non-radiating sources, known as anapoles

(from Greek "ana," "without," thus meaning "without poles"), has recently made possible thanks to the opportunities provided by the multipolar response of high refractive index engineered dielectric nanoparticles known as meta-atoms [3]. In particular, contrary to the plasmonics interactions, these subwavelength scatterers and their two-dimensional (2D) periodic arrangements, known as metasurfaces [4], have been shown to provide an alternative route to manipulate light through the excitation of different cavities, or Mie resonances [5], and enabled several exotic phenomena and applications such as Kerker effects [6], beam steering [7], holography [8] and nonlinear harmonic generation [9].

The theory of electromagnetic multipole expansion, including charge-current spherical and Cartesian decompositions, underpin the light-matter interaction at the nanoscale and provide an opportunity to introduce what so-called toroidal moments family and unambiguously distinguish them from their primitive counterparts, enabling a platform to achieve unprecedented optical phenomena such as the formation of non-scattering anapole [10-12] and super-scattering [13-15] states. In particular, each multipole in the scattering cross-section (SCR) spectra represents an independent scattering channel that has an upper bound, known as the single-channel limit, which is bounded to $(2l + 1)\lambda^2/2\pi$, wherein $l$ represents the order of the multipole. However, the constructive overlap of at least two resonant modes can potentially surpass such a limit, a phenomenon known as super-scattering, and have shown to enable a plethora of applications such as photovoltaics, solar cells, sensing, and biomedicine [13-15]. Contrary to the super-scattering states, anapoles form upon the destructive interference between primitive moments and their toroidal counterparts, leading to vanishing scattering accompanied by strong energy confinement within the subwavelength scatterers [10]. Owing to such an exotic feature, meta-atoms supporting electric dipole anapole (EDA) states have received considerable attention in diverse applications, including strong exciton coupling [16], second and third harmonic generation [17], [18], Raman scattering [19], guiding energy [20], and lasing [21]. While to date, most of the works in this field of research are limited to the study of EDAs, other types of nonradiative states, such as magnetic dipole anapole (MDA) and electric and magnetic quadrupole anapoles (EQA and MQA, respectively) have not been fully investigated. In particular, the existence of these higher-order anapole states is also possible and can provide enhanced radiation suppression and confinement of electromagnetic energy if they are spectrally overlapped with their dipolar moments [22]. Therefore, non-scattering and super-scattering states represent two opposite extreme scattering regimes, and their implementation in a single meta-atom provides a pathway for nanoscale tunable optical switches.

Despite the recent efforts in generating hybrid anapoles (HAs) [22-26], the precise control of the induced anapoles, regardless of their types, has remained challenging. Here, we demonstrate the design of a polysilicon cuboid meta-atom that supports such a new non-radiative hybrid anapole state formed due to the destructive interference between electric and magnetic primitive and toroidal multipoles. In particular, we show that the presented meta-atom can support higher-order non-radiative states up to electric quadrupole anapole state at the same operating wavelength, leading to significant suppression of the far-field radiation and enhancement of electromagnetic energy inside the subwavelength meta-atom as it is schematically shown in **Figure 1**. We also demonstrate that depending on the orientation of the scatterer, its optical response can be tuned from non-radiative to radiative resonant modes. Moreover, we reveal that breaking the out-of-plane inversion symmetry of the meta-atom from a cuboid to a pyramid provides an opportunity for achieving a strong super-scattering optical response. The presented concept in this paper can open new prospects for various applications, such as enhancing nonlinear conversion efficiency, remote sensing, and communications in scattering media.

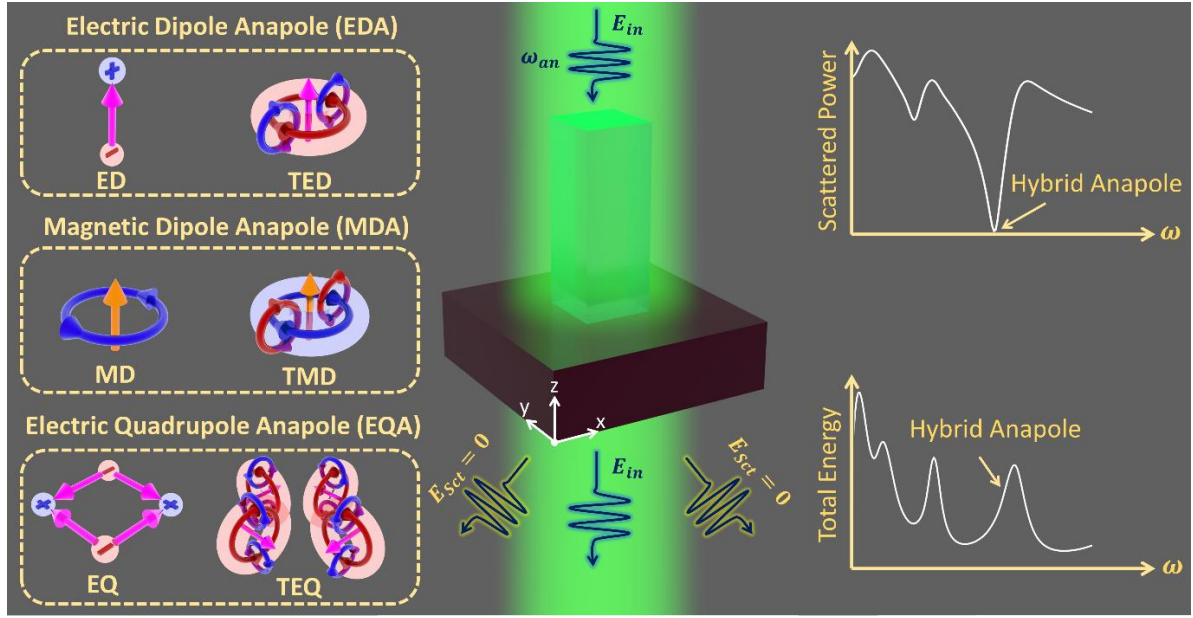

**Figure 1.** The schematic representation of the formation of a hybrid anapole state within an all-dielectric cuboid meta-atom with the width and height of $W = 295$ nm and $H = 466$ nm, respectively. A monochromatic plane wave that is polarized along the $x$-axis normally impinges on the meta-atom and excites nontrivial current configurations, leading to the suppression of the far-field scattering that is accompanied by significant energy confinement. Inset depicts the basic and toroidal moments up to electric quadrupole term with red and blue colors demonstrating the magnetic and electric field currents, respectively.

- **RESULTS AND DISCUSSION**

**Underlying physics of hybrid anapoles.** The optical response of nonmagnetic meta-atoms is characterized based on the multipole expansion approach, which can be performed in two representations of spherical and Cartesian basis [22]. The former is based on the multipole decomposition of electromagnetic fields in terms of the spherical harmonic coefficients, leading the total scattering cross-section to be defined as a series of coefficients as $\sigma_{Sct}(\omega) = \frac{\pi}{k^2}\sum_{l=1}^{\infty}\sum_{m=-l}^{+l}(2l+1)[|a_E(l,m;\omega)|^2 + |a_M(l,m;\omega)|^2]$ where $a_E(l,m;\omega)$ and $a_M(l,m;\omega)$ represent the spherical electric and magnetic multipole scattering coefficients, respectively. On the other hand, the second approach is based on the multipole decomposition of the Cartesian components of induced current density within the meta-atom as $\boldsymbol{J}_{in}(\boldsymbol{r};\omega) = \sum_{l=0}\frac{(-1)^l}{l!}F^{(l)}_{i\ldots k}\partial_i\ldots\partial_k\delta(r)$, with $F^{(l)}_{i\ldots k} = \int_{V_p}\boldsymbol{J}_{in}(\boldsymbol{r};\omega)\,\boldsymbol{r}_i\ldots\boldsymbol{r}_k d\boldsymbol{r}$ denoting a tensor of rank $l$ that corresponds to different Cartesian multipoles [10]. According to the theory of irreducible multipoles, the scattering cross-section of a subwavelength meta-atom can be expressed as [23](see **Methods**):

$$\sigma_{sct}(\omega) = \frac{k_0^4 \sqrt{\epsilon_d}}{12\pi\epsilon_0^2 \mu_0 c} \left| \boldsymbol{p} + \frac{ik_0\epsilon_d}{c} \boldsymbol{T}_1^{(e)} + \frac{ik_0^3 \epsilon_d^2}{c} \boldsymbol{T}_2^{(e)} \right|^2 + \frac{k_0^4 (\epsilon_d)^{3/2}}{12\pi\epsilon_0 c} \left| \boldsymbol{m} + \frac{ik_0\epsilon_d}{c} \boldsymbol{T}^{(m)} \right|^2$$
$$+ \frac{k_0^6 (\epsilon_d)^{3/2}}{160\pi\epsilon_0^2 \mu_0 c} \sum_{u,v} \left| \hat{Q}_{u,v}^{(e)} + \frac{ik_0\epsilon_d}{c} \hat{T}_{u,v}^{(Q_e)} \right|^2 + \frac{k_0^6 (\epsilon_d)^{5/2}}{160\pi\epsilon_0 c} \sum_{u,v} \left| \hat{Q}_{u,v}^{(m)} + \frac{ik_0\epsilon_d}{c} \hat{T}_{u,v}^{(Q_m)} \right|^2 \quad (1)$$

where $\boldsymbol{p}(\omega), \boldsymbol{m}(\omega), \hat{Q}^{(e)}(\omega)$ and $\hat{Q}^{(m)}(\omega)$ correspond to the primitive multipolar moments of electric and magnetic dipoles and quadrupole tensors (ED, MD, EQ and MQ, respectively), while $\boldsymbol{T}_1^{(e)}(\omega)$ and $\boldsymbol{T}_2^{(e)}(\omega)$ denote the electric toroidal dipoles (ETD) of first and second kind, $\boldsymbol{T}^{(m)}(\omega)$ is the toroidal magnetic dipole (TMD), and $\hat{T}^{(Q_e)}(\omega)$ and $\hat{T}^{(Q_m)}(\omega)$ represent the toroidal electric and magnetic quadrupole tensors (TEQ and TMQ, respectively); $c$ is the speed of light in vacuum, $\epsilon_d$ is the dielectric constant surrounding medium, $\epsilon_0, \mu_0$ and $k_0$ are the permittivity, permeability and wavenumber in free space and $u$, and $v$ represent the different components of each tensor. Each of the component of the basic Cartesian multipole moments in Equation (1) can be expressed in terms of the induced current within the particle as

$$p_u(\omega) = \frac{i}{\omega} \int_{V_p} J_{in,u}(\boldsymbol{r}';\omega) \, d\boldsymbol{r}'$$
$$m_u(\omega) = \frac{1}{2} \int_{V_p} (\boldsymbol{r}' \times \boldsymbol{J}_{in}(\boldsymbol{r}';\omega))_u \, d\boldsymbol{r}'$$
$$\hat{Q}_{u,v}^{(e)}(\omega) = \frac{i}{\omega} \int_{V_p} r_u' J_{in,v}(\boldsymbol{r}';\omega) + r_v' J_{in,u}(\boldsymbol{r}';\omega) - \frac{2}{3} \delta_{uv} (\boldsymbol{r}' \cdot \boldsymbol{J}_{in}(\boldsymbol{r}';\omega)) \, d\boldsymbol{r}' \quad (2)$$
$$\hat{Q}_{u,v}^{(m)}(\omega) = \frac{1}{3} \int_{V_p} \left[ (\boldsymbol{r}' \times \boldsymbol{J}_{in}(\boldsymbol{r}';\omega))_u r_v' + (\boldsymbol{r}' \times \boldsymbol{J}_{in}(\boldsymbol{r}';\omega))_v r_u' \right] d\boldsymbol{r}'$$

where $V_p$ is the volume of the meta-atom, the operators of $\cdot$ and $\times$ represent the scalar and vector products, respectively and $\delta_{uv}$ denotes the kronecker delta function. In addition to the primitive multipoles, the contribution of the toroidal moments can also be expressed in terms of induced current as

$$T_{1,u}^{(e)}(\omega) = \frac{1}{10} \int_{V_p} (\boldsymbol{r}' \cdot \boldsymbol{J}_{in}(\boldsymbol{r}';\omega)) r_u' - 2|\boldsymbol{r}'|^2 J_{in,u}(\boldsymbol{r}';\omega) \, d\boldsymbol{r}'$$
$$T_{2,u}^{(e)}(\omega) = \frac{1}{280} \int_{V_p} 3|\boldsymbol{r}'|^4 J_{in,u}(\boldsymbol{r}';\omega) - 2|\boldsymbol{r}'|^2 (\boldsymbol{r}' \cdot \boldsymbol{J}_{in}(\boldsymbol{r}';\omega)) r_u' \, d\boldsymbol{r}'$$
$$T_u^{(m)}(\omega) = \frac{i\omega}{20} \int_{V_p} |\boldsymbol{r}'|^2 (\boldsymbol{r}' \times \boldsymbol{J}_{in}(\boldsymbol{r}';\omega))_u \, d\boldsymbol{r}' \quad (3)$$
$$\hat{T}_{uv}^{(Q_e)}(\omega) = \frac{1}{42} \int_{V_p} \left[ 4 (\boldsymbol{r}' \cdot \boldsymbol{J}_{in}(\boldsymbol{r}';\omega)) r_u' r_v' + 2 (\boldsymbol{r}' \cdot \boldsymbol{J}_{in}(\boldsymbol{r}';\omega)) |\boldsymbol{r}'|^2 \delta_{uv} \right.$$
$$\left. - 5 \left( r_u' J_{in,v}(\boldsymbol{r}';\omega) + r_v' J_{in,u}(\boldsymbol{r}';\omega) \right) |\boldsymbol{r}'|^2 \right] d\boldsymbol{r}'$$
$$\hat{T}_{uv}^{(Q_m)}(\omega) = \frac{i\omega}{42} \int_{V_p} |\boldsymbol{r}'|^2 \left[ (\boldsymbol{r}' \times \boldsymbol{J}_{in}(\boldsymbol{r}';\omega))_u r_v' + (\boldsymbol{r}' \times \boldsymbol{J}_{in}(\boldsymbol{r}';\omega))_v r_u' \right] d\boldsymbol{r}'$$

As can be seen from Equations (1-3), the direct consequence of irreducible Cartesian multipole expansion is the ability to distinguish higher-order toroidal moments within the SCR spectrum, thus, providing an opportunity to induce various anapole states upon satisfying the following conditions:

EDA: $\boldsymbol{p}(\omega) + \frac{ik_0\epsilon_d}{c} \boldsymbol{T}_1^{(e)}(\omega) = 0$

MDA: $\boldsymbol{m}(\omega) + \frac{ik_0\epsilon_d}{c} \boldsymbol{T}^{(m)}(\omega) = 0$

EQA: $\hat{Q}^{(e)}(\omega) + \frac{ik_0\epsilon_d}{c} \hat{T}^{(Q_e)}(\omega) = 0$

MQA: $\hat{Q}^{(m)}(\omega) + \frac{ik_0\epsilon_d}{c} \hat{T}^{(Q_m)}(\omega) = 0$

(4)

The provided expressions of Equation (4) indicate the conditions for exciting any type of higher-order anapole state (up to MQA) at a particular spectral position of $\omega = \omega_{\text{an}}$. It should be remarked that while the destructive interference between the first and second type of electric toroidal dipole moments ($\boldsymbol{T}_1^{(e)}(\omega) + ik_0^2\epsilon_d \boldsymbol{T}_2^{(e)}(\omega) = 0$) leads to what so-called toroidal anapoles, on account of negligible strength of $\boldsymbol{T}_2^{(e)}(\omega)$, we will neglect such a regime of light-matter interaction hereafter (see the second section of Supporting Information for more details and numerical simulations).

**Dynamics of hybrid anapole.** To investigate the excitation of HA within a polycrystalline silicon cuboid meta-atom, the numerical simulations are carried out using the finite-element method (FEM) implemented in the commercial software COMSOL Multiphysics (see **Methods** for details). The cuboid scatterer is considered to have a squared cross-section with the width of $W$ in both $x$ and $y$ directions and height of $H$ in the $z$ direction, as shown in Figure. 1. To investigate the effect of the geometry of the cuboid meta-atom on its scattering characteristics, we firstly fixed the meta-atom width to $W = 295$ nm and change its height from 300-500 nm, while in the second scenario the height is set to $H = 466$ nm and the width is changed from 100-310 nm over the wavelength range of 850-1150 nm. The contributions of the various multipolar moments were obtained by integrating the displacement current induced within the meta-atom using Equations (2) and (3) and their corresponding normalized total scattering cross-sections are shown in **Figure. 2(a)** and 2(b) in logarithm scale for the former and latter cases, respectively. As can be seen from these panels, by varying the height and width of the meta-atom, its scattering response varies significantly, yielding the emergence and suppression of new resonant modes for various spectral positions. In particular, when the height and width of the scatterer changes in the range $420$ nm $< H < 500$ nm and $275$ nm $< W < 310$ nm significant suppressions in the optical response of the meta-atom can be observed, marked by dashed lines, which their lowest non-normalized value reach to $\sigma_{\text{Total}}(\omega)|_{\min} \approx 3 \times 10^{-4}$ $\mu m^2$.

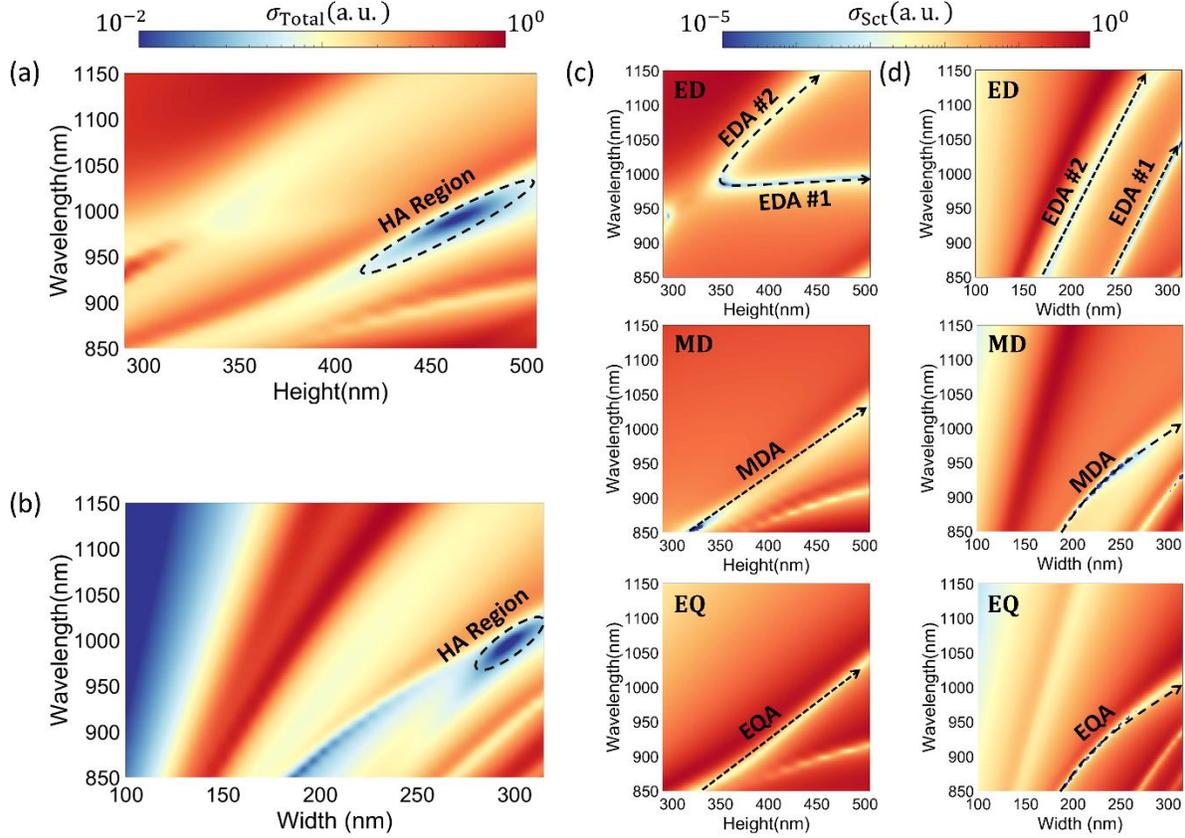

**Figure 2.** The normalized total scattering cross-section of the cuboid meta-atom under the continuous change of its (a) height and (b) width as a function of operating wavelength under plane wave illumination in logarithm scale. The contribution of ED, MD, and EQ multipolar moments excited within the meta-atom as a function of (c) height and (d) width with respect to wavelength. By changing the topology of the meta-atom, several resonances are emerged and suppressed at distinct spectral positions, leading to the formation of resonant branches that redshift toward longer wavelengths as the size of the meta-atom changes.

To clarify the physical origin of such a significant suppression of far-field radiation, the contribution of each higher-order multipolar moment (i.e., primitive + toroidal) is plotted in Figure. 2(c) and 2(d) as functions of height and width over the same wavelength range of 850-1150 nm, respectively. As can be seen from these panels, upon varying the geometrical dimensions of the scatterer, several branches of resonant modes, marked by black dashed lines, emerged within the scattering spectra. Contrary to the optical response of other multipoles, the ED resonant mode possesses two branches of resonances upon varying the dimensions of the meta-atom, which redshift toward longer wavelength as both the height and width of the scatterer approach to larger values. More precisely, changing the height of the meta-atom can significantly shift the emerged dips within the MD and EQ spectra, whereas the ED and resonant modes shift dominantly by varying the width of the nanoparticle (see Supporting Information section III for simulation results of MQ). Therefore, by changing the dimensions of the meta-atom, the spectral position of the dips in its scattering spectra can be overlapped at the same operating wavelength, leading to the regions with suppressed radiation shown in panels (a) and (b).

Despite the existence of spectral regions with suppressed far-field radiation, such an optical response does not necessarily indicate the excitation of the HA state and the satisfaction of conditions given in Equation (4). Indeed, the suppression of the far-field radiation pattern may be attributed to the near-zero values of the induced currents within the meta-atom (i.e., $J_{in}(r,\omega) \approx 0$), which yields the simultaneous suppression of both primitive (Equation (2)) and toroidal (Equation (3)) moments. However, such an optical behavior differs from the HA response as it does not lead to the confinement of energy within the particle [26]. According to Equation (4), to excite an anapole state of any type, its corresponding primitive and toroidal moments should destructively interfere with one another, that is $I_l = |P + T|^2 = |P|^2 + |T|^2 + 2|P||T|\cos(\varphi_P - \varphi_T) \approx 0$, where $I_l$ indicate the intensity of a particular resonant mode, while $P$ ($\varphi_P$) and $T$ ($\varphi_T$) denote the amplitude (phase) of its primitive and toroidal contributions, respectively. In this perspective, two conditions of $|P| \approx |T|$ and $\varphi_P - \varphi_T \approx \pm\pi$ should be satisfied simultaneously, such that the intensity of a particular channel vanishes in the far-field []. To study the origins of the emerged dips within the scattering spectra, we evaluate the explicit contributions of both primitive and toroidal moments as functions of height (400 nm $< H <$ 500 nm) over the wavelength range of 900-1200 nm (the readers are referred to the fourth section of the Supporting Information for the results of the width dependency). For this purpose, we identify the regions wherein the conditions of inducing anapoles $|P| - |T| \approx 0$ and $\varphi_P - \varphi_T \approx \pm\pi$ are satisfied with respect to the operating wavelength and height of the meta-atom as it is shown in **Figure 3**. It should be mentioned that since the incident electric field is polarized along the $x$-direction, we have merely considered the contributing multipoles of $p_x, m_y, \hat{Q}^{(e)}_{xz}$ and $T^{(e)}_{1,x}, T^{(m)}_y, \hat{T}^{(Q_e)}_{xz}$ for primitive and toroidal moments, respectively.

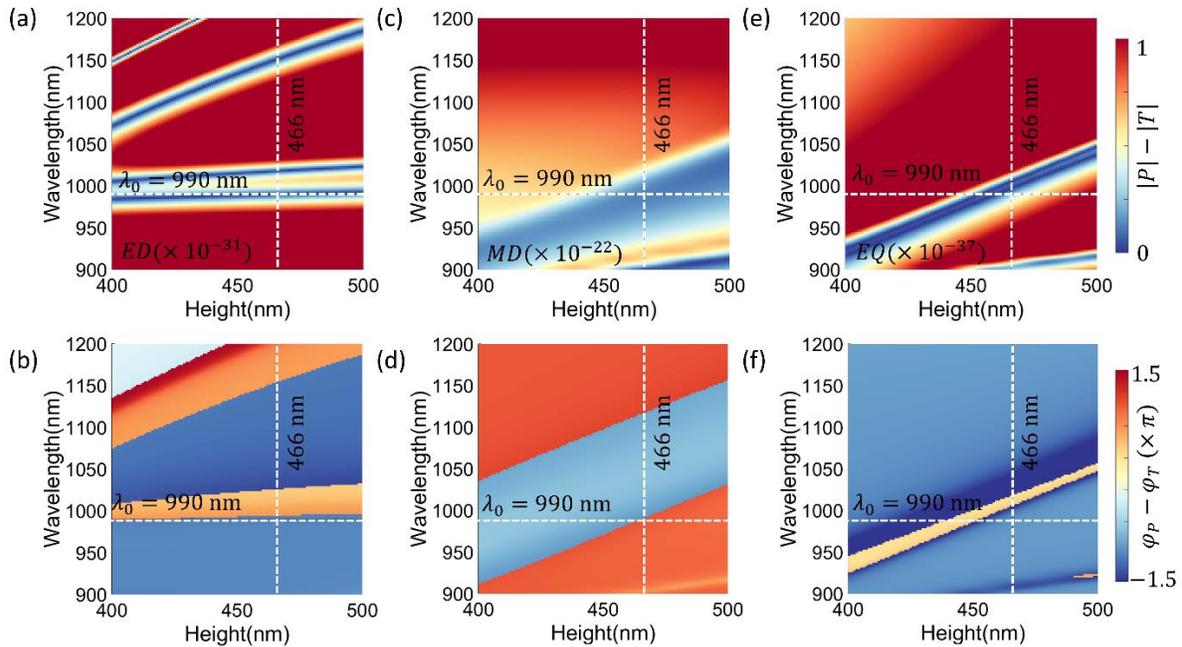

**Figure 3.** The explicit contributions of amplitude and phase of primitive and toroidal moments as functions of height and wavelength for (a,b) ED, (c,d) MD, and (e,f) EQ. The anapole of any kind is excited in regions wherein both conditions are simultaneously satisfied. The vertical dashed line corresponds to the dimension at which the HA state can be excited up to the EQA at the operating wavelength of $\lambda = 990$ nm (horizontal dashed line)

As can be seen from Figures 3(a) and 3(b), while several branches of resonant modes emerge within the amplitude spectra of the ED multipole, the phase difference condition ($\varphi_{p_x} - \varphi_{T_x} \approx \pm\pi$) is merely fulfilled for two branches, leading to the excitation of electric dipole anapoles that are red-shifting toward longer wavelength once the height of the meta-atom increases from 400 nm to 500 nm. On the other hand, despite the fact that individual amplitude and phase conditions for MD moment occur at a broader spectral range (see Figure 3(c) and 3(d)), the magnetic dipole anapole is excited at a smaller region compared to its electrical counterpart, since both conditions are fulfilled simultaneously in a narrow spectral region. Moreover, the same scattering response as that of the EDA can be observed in the results of electric quadrupole anapole, as it is shown in Figures 3(e) and 3(f). In particular, contrary to the recent studies [22,24,25], both the amplitude and phase conditions for exciting ideal EQA are satisfied in the spectral range of interest for the cuboid geometry, making it an appropriate platform for confining electromagnetic energy at the nanoscale. To the best of our knowledge, this is the first time that an ideal electric quadrupole anapole is excited within an all-dielectric subwavelength scatterer. It should be noted that despite the excitation of three higher-order anapoles at the spectral window of interest, the MQA conditions are not satisfied within this region, which is mainly attributed to the fact that the phase difference between $\widehat{M}_{yz}$ and $\widehat{T}_{yz}^{(Q_m)}$ is much smaller than $\pi$. Nevertheless, for $H < 450$ nm the presented meta-atom can support MQA, which thus far is the first theoretical observation of such a peculiar optical behavior (see Supporting Information section III for more details). According to the provided results of Figure 3, we choose the dimensions of the cuboid meta-atom to be $w = 295$ nm and $H = 466$ nm, shown with dashed white vertical line, which can support a hybrid anapole state at $\lambda = 990$ nm (indicated with a horizontal line) up to electric quadrupole anapole. To explicitly demonstrate the formation of HA within the designed scatterer, we calculate its optical response with respect to the operating wavelength, as it is shown in **Figure 4**.

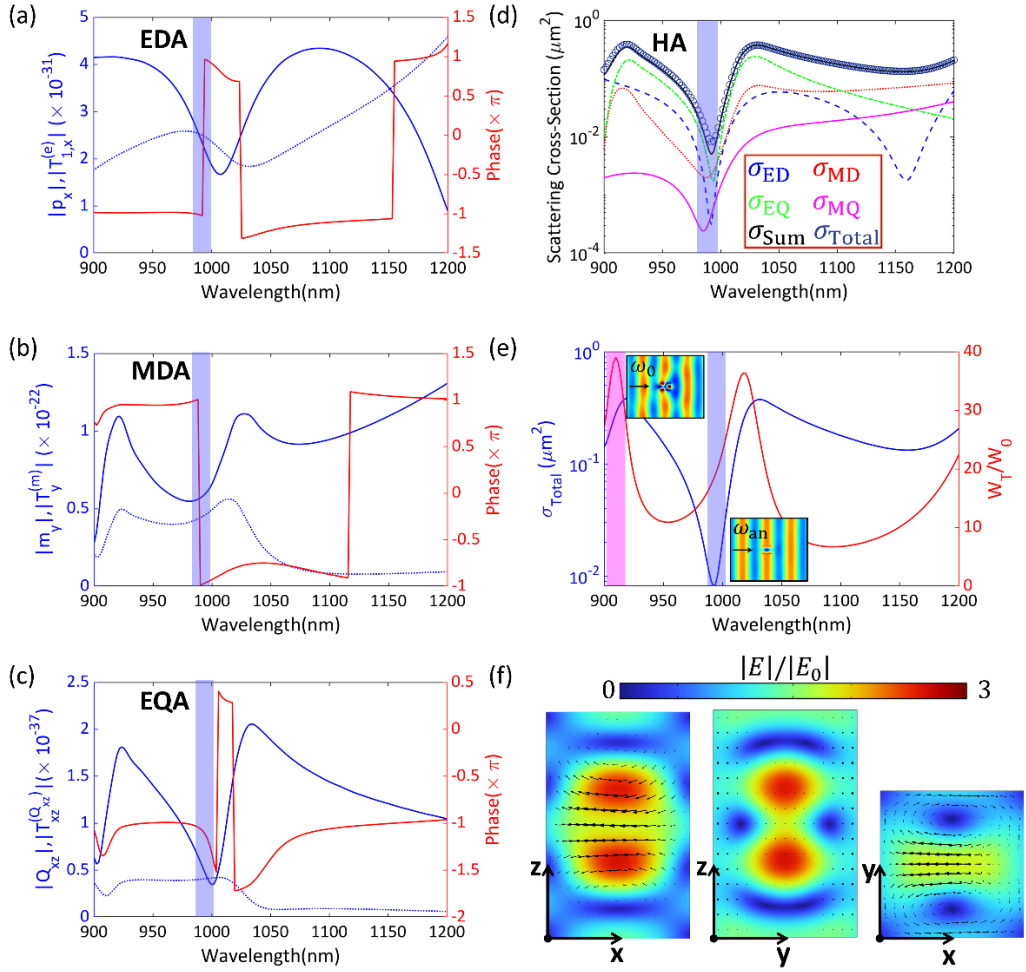

**Figure 4.** The calculated amplitudes and phase differences between the primitive and toroidal multipoles of a cuboid meta-atom with $W = 295$ nm and $H = 466$ nm for (a) ED, (b) MD, and (c) EQ scattering channels. For all the provided plots, the amplitudes correspond to the left ordinate axis with a solid line representing basic moments and dashed curves showing their toroidal counterparts, and the phase differences are read from the right ordinate axis. (d) The multipole decomposition of the cuboid meta-atom up to the magnetic quadrupole contribution. In the legend caption, "Sum" denotes the summation of the contributing moments up to the MQ term, while "total" implies the total scattering cross-section calculated directly from COMSOL. (e) the calculated spectra corresponding to the radiated power (blue color) and the stored electromagnetic energy normalized to the transparent particle (red color). The purple and pink color bands denote the spectral positions corresponding to the dark and bright resonant modes, respectively, with their near-field profiles given at the inset. (f) The electric field distribution at the operating wavelength of 990 nm in three different planes of (z-x), (z-y), and (y-x), with black arrows representing the displacement currents.

The provided results in Figure 4(a-c) demonstrate the amplitude of the primitive (solid line) and toroidal (dotted curve) moments together with their relative phase difference up to electric quadrupole multipole as a function of the operating wavelength. As can be seen from these panels, at the operating wavelength of $\lambda = 990$ nm, the general conditions of Equation (4) are satisfied ( $|P| \approx |T|$ and $\angle \varphi_P - \angle \varphi_T \approx \pm\pi$), leading to the formation of a hybrid anapole state within the nanoresonator. In particular, by carefully designing the geometry of the cuboid meta-atom, it is possible to excite the anapoles of all the contributing moments in the close vicinity of each other, leading to a strong suppression in the far-field radiation as is shown in the scattering spectra of the meta-atom in Figure 4(d). It should be remarked that despite the existence of a dip in the spectra of $\sigma_{MQ}$, such a minimal value does not correspond to the formation of MQA as the conditions of the phase are not satisfied at this wavelength (see section III of the Supplementary File for more information). In addition, the sum of the contributing moments ($\sigma_{\text{Sum}}(\omega) = \sigma_{ED} + \sigma_{MD} + \sigma_{EQ} + \sigma_{MQ}$), and the total scattering cross-section directly computed from COMSOL exhibit good agreement with each other, proving that only the first four multipoles (up to MQ) are sufficient to explain the optical response of the cuboid meta-atom. The energy confinement at different spectral positions within the scattering spectra of the scatterer is also demonstrated in Figure 4(e). For this purpose, we have evaluated the total electromagnetic energy, $w_T = \frac{n^2}{2} \iiint |E(\mathbf{r};\omega)|^2 d^3\mathbf{r} + \frac{\mu_0}{2} \iiint |H(\mathbf{r};\omega)|^2 d^3\mathbf{r}$, integrated over the meta-atom volume and normalized it to the trivial case of a transparent particle with the total energy of $w_0$. As it is highlighted with purple color, at the condition of zero scattering, the calculated energy within the particle is non-zero and is almost 20 times larger than that of the trivial case ($w_T/w_0 \approx 20$), which indicates the significant concentration of energy distribution inside the particle. To the best of our knowledge, this is the first time that such a giant confinement of electromagnetic energy is obtained within a single isolated scatterer at the nanoscale. We also note that despite the existence of other higher energy peaks, such as the one highlighted with pink color in Figure 4(e), their corresponding scattering responses are not suppressed, making them visible to the outside detectors, as it is shown in the inset (see Supplementary Movie 1 and 2). However, the simultaneous overlap of the corresponding zeros of the contributing moments leads to the low-scattering regime functioning as an invisible cloak that stores energy, as shown in the inset of panel (e). As was mentioned earlier, while most of the previous works in the field of non-radiating states were limited to the excitation of electric dipole anapoles, recently, the HA state excited within a cylindrical meta-atom has shown to be capable of better suppression of radiation and higher energy confinements compared to EDAs [24]. In this perspective, we have also compared the optical response of our proposed cuboid meta-atom with that of a cylindrical particle that hosts first, second, and third-order electric dipole anapoles (see Supporting Information section V for more details). Despite the significant contrast between the volumes of the two meta-atoms, ($V_{\text{cuboid}}/V_{\text{cylinder}} \approx 3$), the scattering intensity of the HA state is found to be suppressed more than 16 times compared to its first-order EDA counterpart, whereas its stored energy is almost three times higher. Such a counterintuitive optical response contradicts the conventional wisdom in optics, such that larger objects can, in fact, scatter less light compared to smaller objects. We also note that the HA state can suppress the radiation to the far-field 40 times better than its higher-order electric dipole anapoles, while its stored energy exceeds 2.5 of second- and third-order EDAs. Figure 4(f) shows the distribution of the electric field within the cuboid meta-atom at the corresponding wavelength of the HA state in three different planes of (z-x), (z-y), and (y-x). Contrary to the field distribution in the (y-x) plane, which renders the well-known EDA field profile with a vortex-like distribution of the displacement current shown with black arrows, the internal distribution of the fields in (z-x) and (z-y) planes exhibit more complicated profiles that are attributed to the mixture of electric and magnetic multipole moments and points at a different physical origin compared to EDA. As the final remark, we note that contrary to the concept of dark mode in plasmonics, where a resonance with null net dipole moment cannot be excited by a planar incident wave due to symmetry reasons, the provided HA state can be efficiently excited by a plane wave (or Gaussian beam), making it a feasible platform for experimental investigations.

**From Non-Scattering to Super-Scattering.** It has been shown that the meta-atom orientation (equivalent to the change of illumination direction) can change the optical response of subwavelength particles such that new resonant peaks are emerged/suppressed within the scattering spectra [27]. Such an effect is attributed to the change of different components of multipolar moments and the variation in the coupling efficiency of a particular Mie resonant mode once the illumination direction is changed. In this perspective, tuning the illumination angle can be used as another degree of freedom for controlling the spectral position of the higher order anapoles in the spectral window of interest. Here we investigate the effect of such a combination on the optical response of the presented meta-atom and clarify how higher-order anapoles behave once the illumination direction varies. For this purpose, we fixed the dimensions of the cuboid particle to be $W = 295$ nm and $H = 466$ nm and sweep the orientation angle along the $x$-direction from $0° < \theta < 90°$ over the wavelength range of $900$ nm $< \lambda < 1200$ nm.

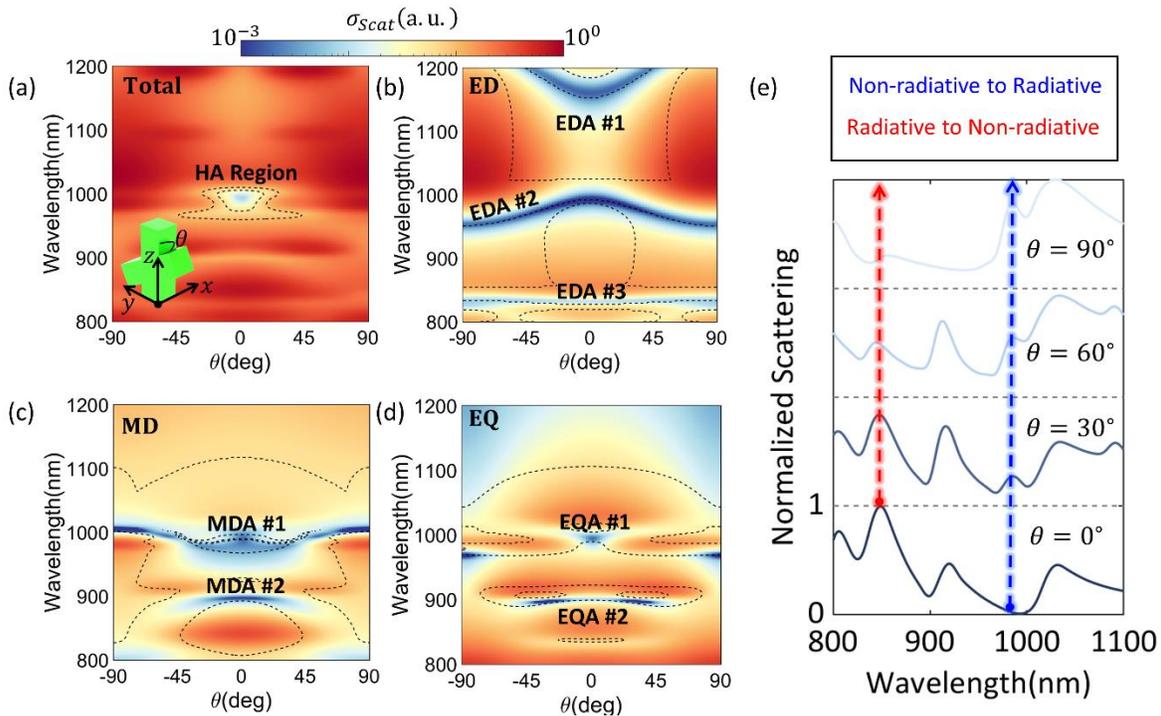

**Figure 5.** (a) The normalized total scattering cross-section of the cuboid meta-atom under plane wave incidence with different angles ranging from $0° < \theta < 90°$ as functions of the operating wavelength. The calculated multipole decomposition results for (b) ED, (c) MD, and (d) EQ, with the dashed line representing the regions wherein $\varphi_P - \varphi_T = \pm\pi$. The anapoles of any kind occur at the points in which the dashed lines coexist with the scattering minima. (e) Total scattering cross-section for four different values of the incident angles $\theta = [0°, 30°, 60°, 90°]$ as functions of wavelength. Tuning the angle of incidence

varies the optical behavior of the meta-atom and changes its response from non-radiative to radiative (blue line) and vice-versa (red line) at two distinct wavelengths.

**Figure 5(a)** shows the total scattering cross-section of the cuboid meta-atom, which exhibits a symmetric response with respect to the angle of incidence due to the symmetrical topology of the scatterer in the plane of incidence. As can be seen from this figure, the illumination direction yields the emergence and suppression of new resonant peaks leading to the formation of a region wherein the radiation to the far-field are strongly suppressed (HA state). More importantly, as opposed to the conventional EDAs, the excitation of such a non-radiating state is not strictly dependent on the frontal illumination (i.e., $\theta = 0°$), and a wide range of incident angles ranging from $-45° < \theta < 45°$ can be employed to achieve such a unique regime of light-matter interaction as it is shown with black dashed line in Figure 5(a). The multipole decomposition results shown in Figures 5(b-c) demonstrate the origin of HA excitation upon changing the angle of incidence. As can be seen from these figures, tuning the angle of incidence yields the excitation/suppression of new resonant modes, revealing another mechanism for engineering the Mie-type resonances within the meta-atom. In particular, several dips in the scattering spectra of the meta-atom are observed under oblique incidences, which are attributed to the destructive interference between the primitive and toroidal moments of the contributing Mie resonances as long as they are overlapping with the black dashed lines that are representing the regions wherein the phase difference conditions for inducing anapole states ($\varphi_P - \varphi_T \approx \pm\pi$) are satisfied. Taking the electric dipole response as the example, we note that merely three lines coexist with the emerged resonant dips, indicating that for these particular branches, both the amplitude and phase conditions are satisfied, and EDA can be obtained and tuned with respect to the incident angle. For the magnetic dipole and electric quadrupole anapoles, such a behavior becomes more complicated, and the anapole conditions are satisfied in two regions shown in the plots. Interestingly, the higher order anapole states can also be obtained for lateral illuminations ($\theta = 90°$), indicating that the proposed meta-atom can be used for on-chip photonics wherein the resonant behavior and energy confinement from side illumination is highly demanded. The provided results in Figure 5(e) further elaborate on how the optical response of the meta-atom changes once the incident angle varies from frontal ($\theta = 0°$) to lateral ($\theta = 90°$) illumination in the steps of 30°. While the overall scattering response of the scatterer change significantly once the illumination angle varies, the optical behavior of the meta-atom at the operating wavelength of $\lambda = 990$ nm gradually alters from a non-radiating hybrid anapole state (for $\theta = 0°$) to radiative resonant mode corresponding to a magnetic dipole (for $\theta = 90°$) as it is shown with blue dashed curve. Contrary to such an optical behavior, the scattering response of the presented meta-atom can be tuned at $\lambda = 850$ nm from radiative state to electric dipole anapole upon the continuous change of illumination angle from $\theta = 0°$ to $\theta = 90°$, respectively. We note that the observed optical phenomena are not related to the common shifts in the scattering spectra studied previously and are directly attributed to the change in the coupling efficiency of a particular Mie resonant mode once the direction of illumination is changed.

So far, we have shown the excitation of a peculiar optical response of HA within a cuboid meta-atom possessing inversion symmetry along the -axis. In particular, the inversion symmetry of the optical scatterer forces the eigenmodes to be consisted of either merely even (MD, EQ, etc.) or odd (ED, MQ, etc.) multipoles, which can constructively/destructively interfere with one another as it is schematically shown in **Figure 6(a)**. However, as it has been recently shown by Poleva et al. [28], breaking the inversion symmetry of the particle can lead to bianisotropic responses such that the ED and MD are mutually coupled, and the corresponding eigenmodes of the meta-atom become a combination of multipoles with mixed parities (see Figure 6(b)), leading to the amelioration of the directivity of a particular eigenmode. Here, we demonstrate that by breaking the inversion symmetry of the meta-atom from a cuboid to a pyramid geometry, the scattering response is changed from a non-scattering hybrid anapole state (for cuboid geometry) to a super-scattering regime (pyramid structure) as a result of the transition from non-bianisotropic to bianisotropic optical responses. In particular, each multipole in the scattering cross-section spectra represents an independent scattering channel that has an upper bound, known as the single-channel limit, which is bounded to $(2l + 1)\lambda^2/2\pi$, wherein $l$ represents the order of the multipole [13-15]. Conventionally, the notation of super-scattering applies to the cases where the total scattering cross-section surpasses the given single-channel limit of an electric dipole ($l = 1$) such that $\frac{\sigma_{Total}}{\sigma_{ED}^{max}} = \frac{2\pi\,\sigma_{Total}}{3\lambda^2} > 1$ [29]. To clarify how the transition of the topology of the meta-atom from non-bianisotropic to bianisotropic can lead to a super-scattering regime, we fixed the height of the scatterer to be the same as that of previous sections while the top-to-bottom width ratio ($\beta = W_T/W_B$) is changed in the range of $0.1 < \beta = \frac{W_T}{W_B} < 1$ and calculated $\frac{2\pi\,\sigma_{Total}}{3\lambda^2}$ with respect to $\beta$ and operating wavelength as it is shown in Figure 6(c).

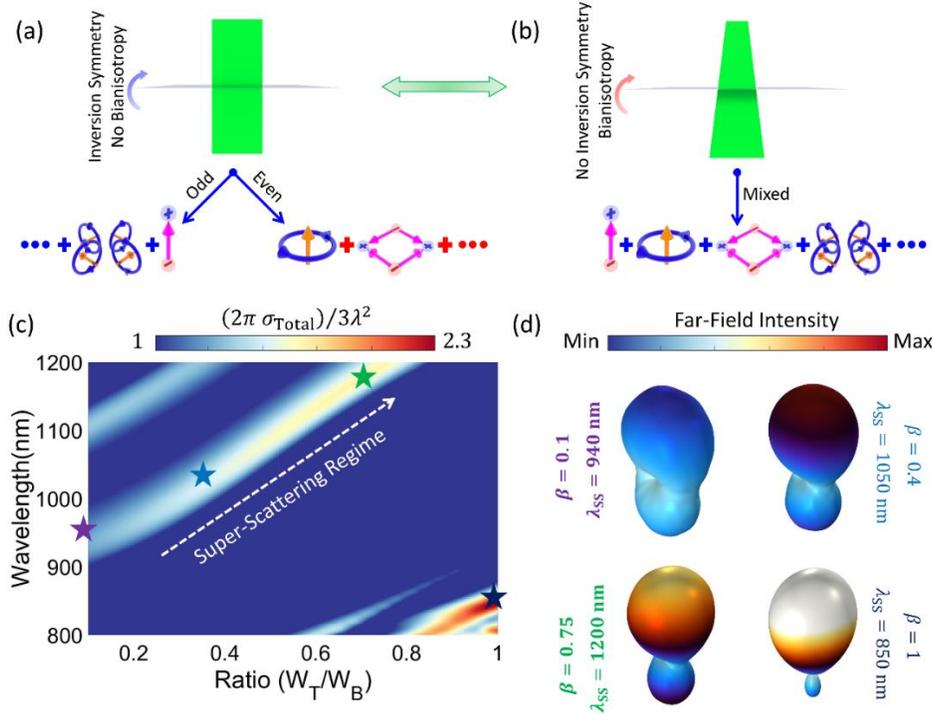

**Figure 6.** The schematic demonstration of achieving bianisotropic responses via breaking the inversion symmetry of the meta-atom. For (a) symmetric particle, the eigenmodes consist of either only even or odd multipoles, while for (b) asymmetric meta-atom, the optical response consists of multipoles of mixed parities. (c) The calculated response of the pyramid meta-atom with respect to the top-to-bottom width ratio and operating wavelength. The supper-scattering regime occurs once $\frac{2\pi \, \sigma_{Total}}{3\lambda^2} > 1$. (d) The radiation pattern of the far-field intensity calculated at the wavelength corresponding to the super-scattering state for the meta-atoms with dimensions shown in panel (c).

As can be seen from the calculated response, strong super-scattering regime ($\frac{2\pi \, \sigma_{Total}}{3\lambda^2} \approx 2.3$) can be obtained for a near perfect inversion symmetric meta-atom ($0.8 < \beta < 1$) in a narrow spectral region, whereas an asymmetric meta-atom ($\beta < 0.8$) can support supper-scattering optical response in a broad spectrum of 900 nm-1200 nm with lower amplitude of 1.6. Such an optical response is a direct result of breaking inversion symmetry which not only leads to bianisotropic responses but also achieves super-scattering behavior that exceeds twice of single-channel limit. The contributions of each moment (up to MQ) are also calculated and provided in section VI of the Supporting Information. In Figure 6(d), we demonstrate the far-field intensity of the meta-atom for $\beta = [0.1, 0.4, 0.75, 1]$ at the operating wavelength corresponding to the maximal value of the super-scattering spectrum marked in panel (c). As can be seen from these radiation patterns, while for all the proposed cases directional scattering can be observed on account of establishing super-scattering regime, the magnitudes of the radiation patterns increases once the meta-atom geometry changes from asymmetric to inversion symmetric topology. Such a variation in the directivity of the meta-atoms is attributed to spectral overlap of all the resonant modes for $\beta = 1$, whereas for other cases ($\beta = [0.1, 0.4, 0.7]$) fewer multipoles are overlapping with one another. These results demonstrate the role inversion symmetry as another tuning mechanism to achieve a thoroughly opposite regime of light-matter interaction compared to non-scattering hybrid anapole state.

## Conclusion

In this work, we demonstrated the design of a cuboid meta-atom that supports new non-radiative states, called hybrid anapoles, formed due to the destructive interference between electric and magnetic basic and toroidal multipoles. In particular, we theoretically predicted and numerically showed that such a meta-atom can support higher-order non-radiative states up to electric quadrupole moment at the same operating wavelength, leading to significant suppression of the far-field radiation and enhancement of electromagnetic energy inside the subwavelength dielectric particle. We also explored for the first time the role of illumination angle on the optical response of such a hybrid dark state and demonstrated that depending on the orientation of the meta-atom, its scattering response switch from non-scattering to radiating states of various kinds. Moreover, we revealed that tuning the topology of the meta-atom from a cuboid to a pyramid leads to the formation of so-called super-scattering, which is an opposite regime of light-matter interaction compared to the HA state. The presented concept of hybrid anapole state has several potential applications ranging from remote sensing relying on the shape and orientation of meta-atoms to the enhancement of nonlinear conversion efficiency and establishing strong coupling between photonic and excitonic platforms.

**Methods**

*Numerical Simulations:* The numerical simulations are carried out using the finite-element method (FEM) implemented in the commercial software COMSOL Multiphysics. In particular, we utilize the Wave Optics Module to solve Maxwell's equations in the frequency domain together with proper boundary conditions. Here, we use a spherical domain filled with air and a radius of $4\lambda$ as the background medium, while perfectly matched layers of thickness $0.6\lambda$ are positioned outside of the background medium to act as absorbers and avoid undesired scattering. Tetrahedral mesh is also chosen to ensure the accuracy of the results and allow numerical convergence. As we have recently shown in [], although the type of illumination beam can significantly affect the excited multipolar moments, the key point in the evaluation of scattering cross-section is the calculation of the induced currents within the resonator and then associating the multipolar moments according to the given expressions of Equations (2) and (3). Since in this paper, we are mainly interested in the effect of meta-atom topology in inducing two opposite regimes of light-matter interactions (i.e., HA and super-scattering), we have limited our study to the case of plane wave illumination (electric field polarized along the $x$-axis), yet the interaction of other light beams, known as structured lights, is expected to provide exotic optical responses, which is not in the scope of this work.

*Multipole Decomposition:* Recently, several theoretical frameworks, which are based on the irreducibility of Cartesian tensors with respect to the SO(3) group, have been proposed to unambiguously describe the contributions of toroidal type multipoles in the scattering spectra of optical scatterers [23]. According to the theory of irreducible multipoles, the dyadic-like form of scattered field in the far-field region up to magnetic quadrupole can be expressed as (see Supporting Information section I for more general form)

$$\begin{aligned}\boldsymbol{E}_{Sct}(\boldsymbol{n},\omega) = \frac{k_0^2 \exp(ik_0 r)}{4\pi\epsilon_0 r} &\left( \left[\boldsymbol{n} \times \left\{\left(\boldsymbol{p}(\omega) + \frac{ik_0\epsilon_d}{c}\boldsymbol{T}_1^{(e)}(\omega) + \frac{ik_0^3\epsilon_d^2}{c}\boldsymbol{T}_2^{(e)}(\omega)\right) \times \boldsymbol{n}\right\}\right] \right.\\ &+ \frac{1}{c}\left[\left(\boldsymbol{m}(\omega) + \frac{ik_0\epsilon_d}{c}\boldsymbol{T}^{(m)}(\omega)\right) \times \boldsymbol{n}\right] \\ &+ \frac{ik_0}{2}\left[\boldsymbol{n} \times \left\{\boldsymbol{n} \times \left(\left(\hat{Q}^{(e)}(\omega) + \frac{ik_0\epsilon_d}{c}\hat{T}^{(Q_e)}(\omega)\right) \cdot \boldsymbol{n}\right)\right\}\right] \\ &+ \left. \frac{ik_0}{2c}\left[\boldsymbol{n} \times \left\{\boldsymbol{n} \times \left(\left(\hat{Q}^{(m)}(\omega) + \frac{ik_0\epsilon_d}{c}\hat{T}^{(Q_m)}(\omega)\right) \cdot \boldsymbol{n}\right)\right\}\right] \right) \end{aligned} \quad (5)$$

where $\boldsymbol{n} = \boldsymbol{r}/r$ is the unit vector directed from the particle's center towards an observation point. Using these notations, the far-field scattered power can be readily related to the scattered fields of Equation (5) with the aid of a time-averaged Poynting vector as $dP_{Sct} = 0.5\sqrt{\epsilon_0/\mu_0}|E_{Sct}|^2 r^2 d\Omega$, wherein $d\Omega = \sin\theta d\theta d\varphi$ represents the solid angle. Therefore, combining Equation (1) with the given relation of scattered power and performing the integration over the total solid angle, the scattering cross-section, defined as $\sigma_{\text{Sct}} = 2\sqrt{\mu_0/\epsilon_0 \epsilon_d} \cdot \left(\frac{P_{Sct}}{|\boldsymbol{E}_{inc}|^2}\right)$, with $\boldsymbol{E}_{inc}$ representing the incident wave, can be obtained as that of Equation (1).

## Acknowledgments

This paper was supported in part by the Office of Naval Research (ONR) (Grant No. N00014-20-1-2558), National Science Foundation (NSF) (Grant No. 1809518), and Army Research Office Award (Grant No. W911NF1810348).

## Conflict of Interest

The authors declare no conflict of interest.

## Data Availability Statement

The data that support the findings of this study are available from the corresponding author upon reasonable request.